\documentclass[proof]{WileyASNA-v1}

\articletype{Article Type}%

\usepackage[english]{babel}
\usepackage[utf8]{inputenc}

\begin{document}

\title{Modeling changing-look (CL) AGN phenomenon in 1D using accretion disk instabilities}


\author[1,2]{Marzena Śniegowska*}

\author[2]{Mikołaj Grzędzielski}

\author[2]{Bo\.zena Czerny}

\author[2]{Agnieszka Janiuk}

\authormark{Śniegowska \textsc{et al}}

\address[1]{\orgdiv{Nicolaus Copernicus Astronomical Center  (PAN)},  \orgaddress{\state{ul. Bartycka 18, 00-716 Warsaw}, \country{Poland}}}

\address[2]{\orgdiv{Center for Theoretical Physics, Polish Academy of Sciences, Al. Lotnik\'ow 32/46}, \orgaddress{\state{02-668 Warsaw}, \country{Poland}}}

\corres{*Corresponding author. \email{msniegowska@camk.edu.pl}}




\abstract{Apart from regular, low-level stochastic variability, some AGN occasionally show exceptionally large changes in the luminosity, spectral shape, and/or X-ray absorption. The most notable are the changes of the spectral type when the source classified as a Seyfert 1 becomes a Seyfert 2 galaxy or vice versa. Thus a name was coined of 'Changing-Look AGN' (CL AGN). The origin of this phenomenon is still unknown, but for most of the sources, there are strong arguments in favor of the intrinsic changes.

Understanding the nature of such rapid changes is a challenge to the models of black hole accretion flows since the timescales of the changes are much shorter than the standard disk viscous timescales. 
We aim to model the CL AGN phenomenon assuming that the underlying mechanism is the time-dependent evolution of a black hole accretion disk unstable due to the dominant radiation pressure. We use a 1-dimensional, vertically integrated disk model, but we allow for the presence of the hot coronal layer above the disk and the presence of the inner purely hot flow. We focus on the variability timescales and amplitudes, which can be regulated by the action of large-scale magnetic fields, the description of the disk-corona coupling and the presence of an inner optically thin flow, like Advection-Dominated Accretion Flow (ADAF). We compare model predictions 
for the accretion disk around black hole mass 10$^7$M$_{\odot}$. 
}

\jnlcitation{\cname{%
\author{Śniegowska M.}, 
\author{Grzędzielski M.}, 
\author{Czerny B.} , and
\author{Janiuk A.} } (\cyear{2021}), 
\ctitle{Modelling changing-look (CL) AGN phenomenon in 1D using accretion disk instabilities}, \cjournal{Astronomische Nachrichten}, \cvol{???}.
}


\maketitle


\section{INTRODUCTION}\label{sec1}
In the simplest theoretical framework, active galactic nuclei (AGN) are sources in which a supermassive black hole is surrounded by a geometrically thin, optically thick accretion disk \cite{ss73}. 
For some objects, though, modifications to accretion disk, like advection-dominated accretion flow - ADAF or Slim Accretion Disks (for review see \cite{2019Univ....5..131C}), are needed to better represent the observed spectra. AGN also show variability in different timescale, pattern and wavelengths like X-ray \citep{lawrence1993}, optical \citep{sesar2007} and infrared \citep{kozlowski2016} bands. 


Generally, in accretion disk heating processes is mediated by the turbulent magnetic field. Thus the stochastic variability observed in light curves of AGN is not surprising.
However, objects with significant changes in the luminosity and the noticeable (dis)appearance of broad emission lines are less likely to be explained by stochastic variability. 
They are called changing-look (CL) AGN and the discussion about their physical nature is coming under scrutiny in recent years.

In this paper, we focus on accretion disk around supermassive black holes (10$^7$M$_{\odot}$). 
The variations in global accretion rate in this kind of accretion disks are one of the candidates for CL AGN physical explanation.
However, it seems that those objects may be explained by different scenarios and just one explanation for them is not likely. CL AGN differ between each other in the shape of the light curve, timescale of changes, and its amplitude.
For example, changes in some objects (like Mrk 590 \cite{2020MNRAS.491.4615K}) can be explained by the extrinsic matter which causes the dimming of the flux and spectral changes.
Whereas, for others (like in sources investigated by \citep{2019A&A...625A..54H}) the obscuration is ruled out because of no changes in polarisation degree, which should appear if we assume dusty matter is temporary obscuring the source.  Thus, intrinsic changes are in favor to explain it in most cases. Intrinsic changes in accretion disks are rather a broad concept and may include scenarios with tidal disruption events (i.e SDSS J224113-012108 investigated by \cite{2021MNRAS.500L..57Z}), tidal interaction between disks in supermassive black hole binaries \citep{2020wangbon}, the presence and disappearance of the warm corona \citep{noda2018},
changes in the magnetization of accretion disk \cite{2021MNRAS.502L..50S} or magnetic accretion disk-outflows \citep{2021ApJ...916...61F}.
The rapidly increasing number of observations of CL AGN \citep{2021arXiv210607660S, 2020MNRAS.491.4925G, 2016MNRAS.457..389M} and variety of possible scenarios of changes motivates us to explore this field.

\begin{figure}[t]

\centerline{\includegraphics[scale=0.4]{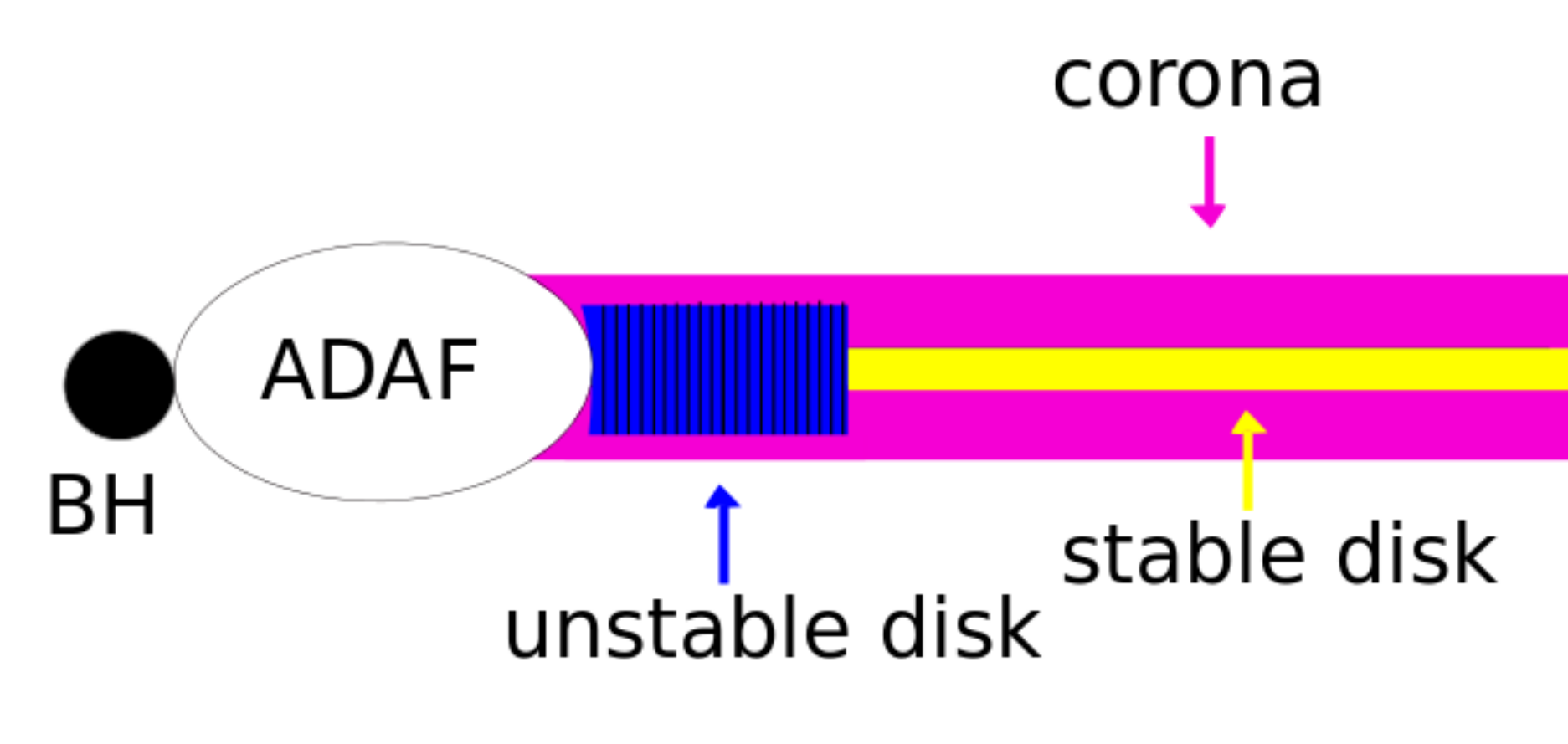}}

\caption{The schematic view of the model sliced through the disk plane: 
from the left black hole marked as a black circle, and inner hot ADAF, disk part, which is unstable due to radiation pressure instability marked in blue, with black stripes to emphasize radial grid, and outer cold stable disk represented as a yellow area. The corona which is covering the disk  is shown in magenta. \label{fig:models}}
\end{figure}

To model CL AGN phenomenon (especially for objects with repetitive outbursts), we assume the radiation pressure instability mechanism.
In the current work, we use time-evolution code GLADIS (Global Accretion Disk Instability Simulation) developed by  \cite{janiuk2002} (recently publicly available version is presented in \cite{2020mbhe.confE..48J}), which allows performing a time-dependent simulation of accreting matter onto massive objects. This code was already used to model the 'heartbeat' outbursts in microquasars \citep[e.g. for IGR J17091-3624 by][]{2015janiuk} as caused by radiation pressure instabilities. The idea of explaining 
the changes in CL AGN with the radiation pressure instability was used in \cite{2020A&A...641A.167S}, however, the concept of the geometry of the system was very simplistic, reducing the unstable region to a single zone both in radial and vertical direction. In this work, we explore the instability using 1-D radially resolved model. 
In Fig. \ref{fig:models} we present the scheme of the model which we use. This model contains the Advection-Dominated Accretion Flow (ADAF) in which hot coronal matter is freely advected inwards,
the unstable standard cold disk zone radially resolved, the stable outer disk and the  geometrically thick, optically  thin corona above the disk. Therefore, in comparison to \cite{2020A&A...641A.167S}, this model is extended by 
two important aspects:
(i) the presence of the disk corona, and (ii) radially resolved unstable cold disk. The grid in the unstable zone is improving the model, because the ring in which instabilities are present is not artificially limited in its size.
Disk instabilities in system disk-corona interacting were investigated for the microquasar GRS 1915+105 by \cite{2005MNRAS.356..205J}, with the disk/corona mass exchange model developed later on \cite{2007A&A...466..793J} for variability in X-ray binaries in their soft state and AGN. The presence of the inner ADAF was not included in these papers.
The structure of the paper is as follows:
In Section \ref{sec2} we introduce the physical assumptions which we use  throughout this work for 3 computed models. In Section \ref{sec3} we show and describe the preliminary cycle-behavior time-dependent model results.
Finally, in Section \ref{sec4}, we conclude on the current results and comment on the plans for future work.
\section{THEORY}\label{sec2}
In the standard accretion disk theory, \citep{ss73} the assumption of the radiation pressure dominance leads to the accretion disk instability (thermal and viscous) which was shown by \cite{1973A&A....29..179P} and \cite{le74}. Those instabilities may appear in the broad range of black hole masses' objects.

Outbursts that may appear due to radiation pressure instabilities in microquasars (masses $\approx$ 10 M$_\odot$) were explored by \cite{janiuk2002}. 
This kind of instabilities in objects with central black hole mass up to 10$^8$ M$_\odot$ was further described by \cite{grzedzielski2017}. The latter work gives also the quantitative prescriptions for estimating periods and amplitudes of outbursts. We use those prescriptions to discuss the predicted limits of the minimum period for black hole mass of 10$^7$ M${_\odot}$.

\subsection{S-curve as the signpost for the initial conditions}

\begin{figure}[t]
\centerline{\includegraphics[scale=0.4]{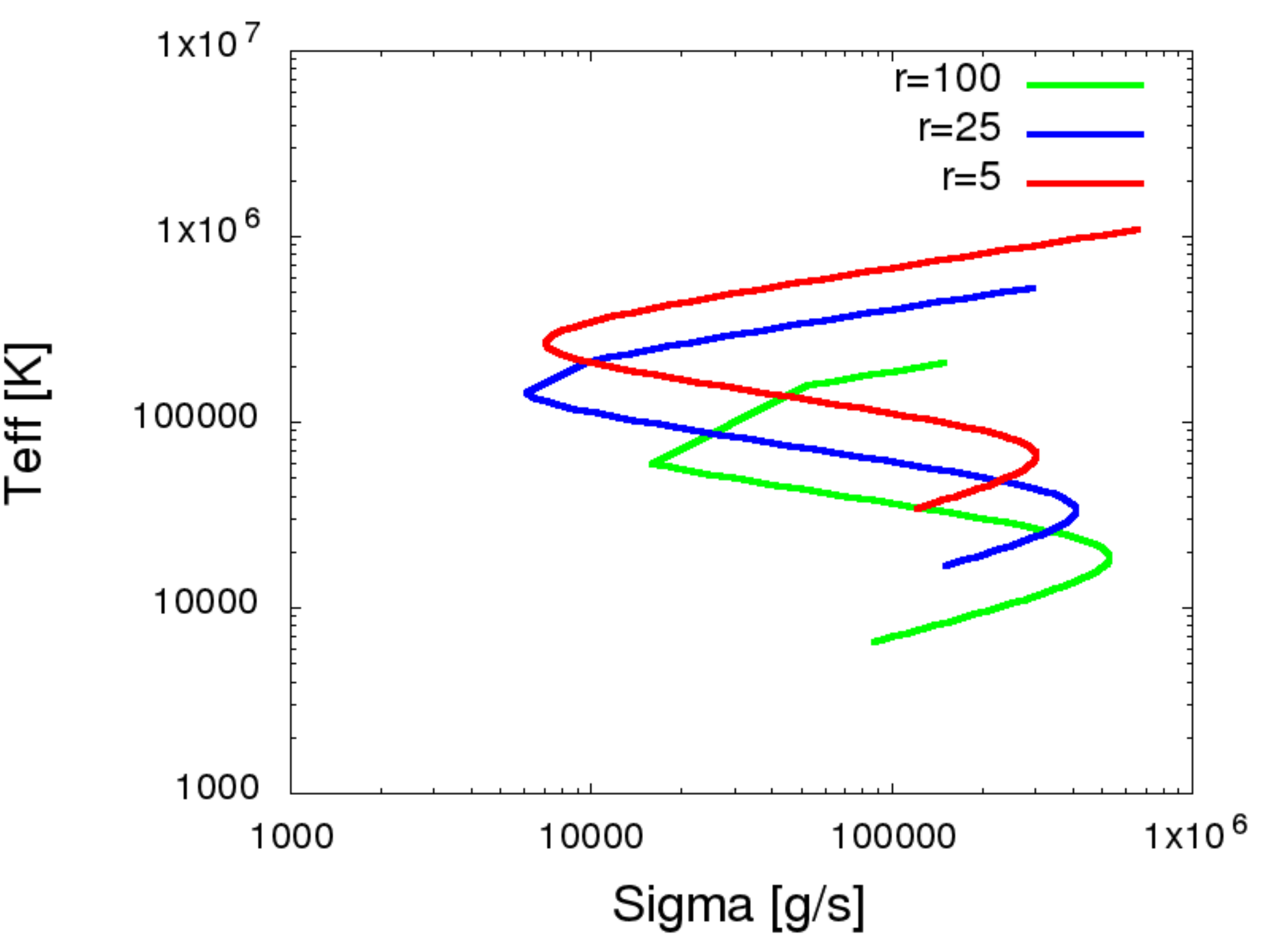}}
\caption{Local stability curves for black hole mass 10$^7 M_\odot$, $\alpha$ = 0.02 and different radii from the top: 5$R_{Schw}$ marked as red curve, 25$R_{Schw}$ marked as blue curve and 100$R_{Schw}$ marked as green curve.\label{scurve}}
\end{figure}

A stability curve, in general, is a sequence
of stationary solutions to equations of disk structure equations, solved locally for surface density and temperature.
Before we perform the time-dependent simulation with the
GLADIS code, we compute the stability S-curves in an effective temperature - surface density plane to get an overall picture (see Fig. \ref{scurve}). 
From the shape of the S-curve, and specifically the range of its unstable branch (i.e., the negative slope of $T-\Sigma$ dependence) we infer the information about the amplitudes and timescales of outbursts. Along the unstable branch, the higher accretion rates correspond to smaller surface densities. The position of the curve depends on the radius.
Hence, with changing the distance from the central source, at a fixed temperature we switch between the stable and unstable branch. 
Note, however, that the local effective temperature scales with the radius and the local accretion rate, hence at a fixed accretion rate the (sufficiently large) outer disk is always located at a stable branch. 
By investigating S-curves behavior before running the time evolution code 
we predict proper initial conditions and model parameters which allow us to obtain instability of a desired range.


\subsection{Set-up of the Model}

\label{itemizetimedependent}
The two time-dependent equations adopted by GLADIS code are equations of hydrodynamics for surface density and temperature evolution under assumptions of energy conservation, viscous diffusion, radiation pressure dominance (see eq. 3.1 and 3.2 in \cite{2020mbhe.confE..48J})


In this work, we investigate 3 possible scenarios of mass exchange and evaporation 
\begin{itemize}
 
    \item  mass exchange between disk and corona is on, and disk evaporation is due to the flux generated in the disk and the corona - case A (see 
    \citealt{2005MNRAS.356..205J}, their Eq.~17, with coefficients $B_1 = 0.5$, $B_2 = 0.5$). This description represents the disk evaporation due to the electron conduction between the disk and the corona,
    \item mass exchange between disk and corona is on, and disk evaporation is due to magnetic field - case B (see  \citealt{2007A&A...466..793J}, their Eq. 10, their case B). This description represents the corona heated by the magnetic buoyancy,
    \item  mass exchange between disk and corona is off, which implies no disk evaporation - case C.
    
\end{itemize}

For all calculated models in this work, we assume that the heating term is proportional to the total pressure, as was assumed by \cite{ss73}.

We fix the following physical parameters throughout all the computed models: the black hole mass 10$^7$M$_{\odot}$, viscosity coefficient $\alpha$ = 0.02 for disk and corona after \cite{grzedzielski2017}, global accretion rate as 0.007 solar masses per year, the inner radius of the disk 20 R$_{Schw}$, the outer radius of the disk 100 R$_{Schw}$\footnotetext{due to limitations in computing time we decided to keep outer radius in this position, however, unstable zones are more extended (see \cite{2011MNRAS.414.2186J})}. We discuss this issue later.
Corona temperature is fixed at the virial temperature, whereas the temperature of the disc evolves in time.
The total duration of each simulation is 274 years in the source time.

\section{RESULTS}\label{sec3}


In this section, we compare the properties of time-dependent models for different underlying assumptions. 
We calculate 3 solutions based on different assumptions about the disk/corona mass exchange (see Sec. \ref{itemizetimedependent}).

\subsection{The outburst cycle}
\begin{figure}[t]
\centerline{\includegraphics[scale=0.5]{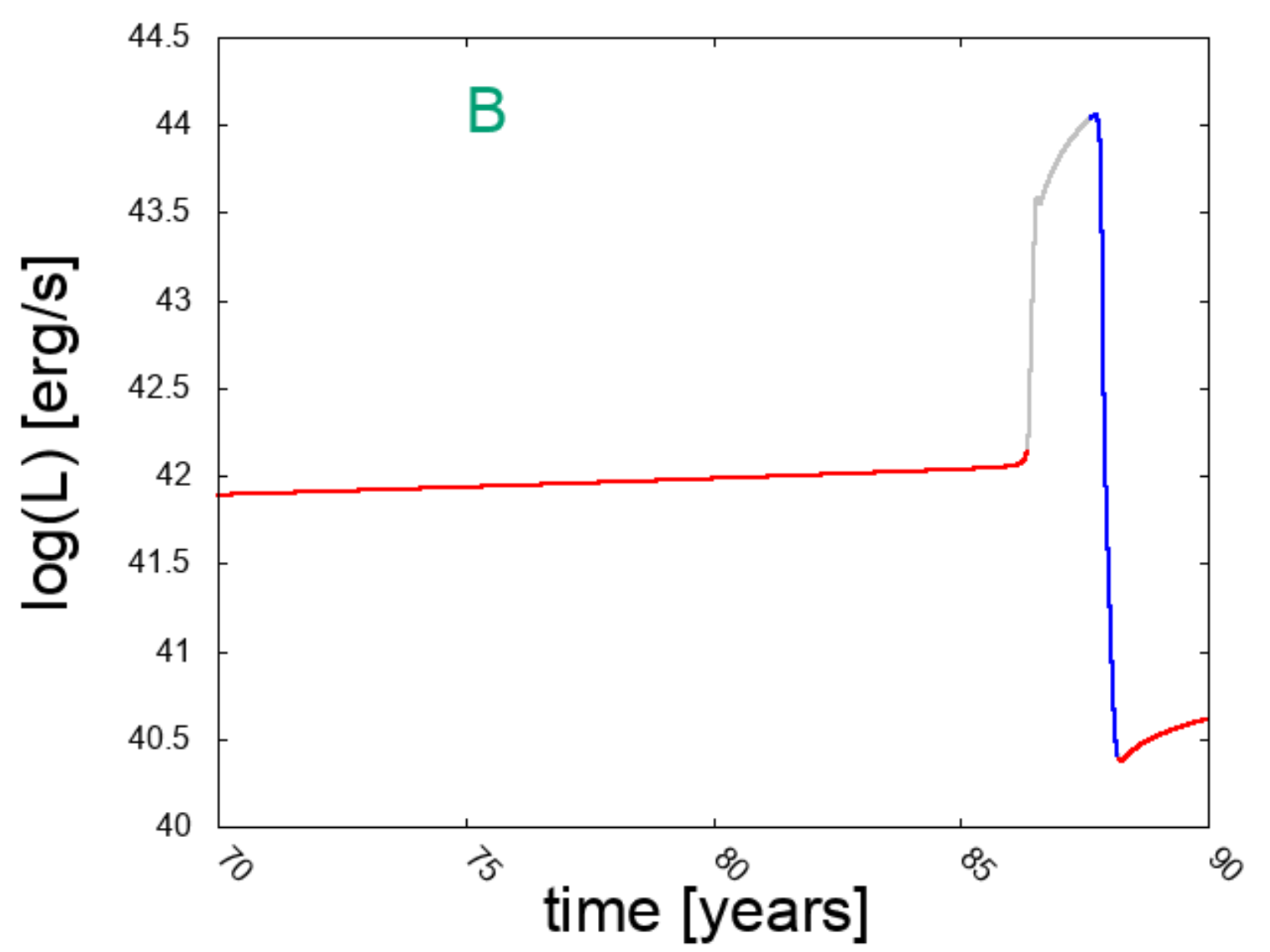}}
\caption{The exemplary fragment of the disk lightcurve with cycle-steps marked for black hole mass 10$^7 M_\odot$, R$_{in}$ = 20R$_{shw}$ and R$_{out}$ = 100R$_{shw}$ for case B. The grey part of the light curve represents heating phase, the blue one advective phase and the red one diffusive phase. Note that we present in this plot just part of the diffusive phase, not the full cycle.  \label{cycle-example}}
\end{figure}

The example of outburst for the object with the black hole mass 10$^7$M$_{\odot}$ is shown in the Fig \ref{cycle-example}.
The luminosities (for disk and for corona) presented in this work are the integrated radiation flux between inner and outer radius.

Each outburst has three characteristic phases:
\begin{enumerate}
    \item viscous phase (red part of the lightcurve), in which the approximate thermal equilibrium holds, and the surface density increases till the system meets the loss of equilibrium (lower branch in the S-curve),
    \item heating phase and hot viscous phase (the grey part of the lightcurve), when the temperature in the inner regions of the disk increases rapidly, and the accretion rate is enhanced, the system then evolves along the upper S-curve branch till the turning point of the S-curve is met,
    \item cooling phase (blue part of the lightcurve), where the temperature and accretion rate drop-down till the lower branch of the S-curve is achieved.

\end{enumerate}
Outbursts can be then characterized by the amplitude, the period, and the relative duration of the outburst in comparison with the total period, as discussed by \citet{grzedzielski2017} for their set of models.

\begin{figure*}[t]
\centering
\includegraphics[scale=0.32]{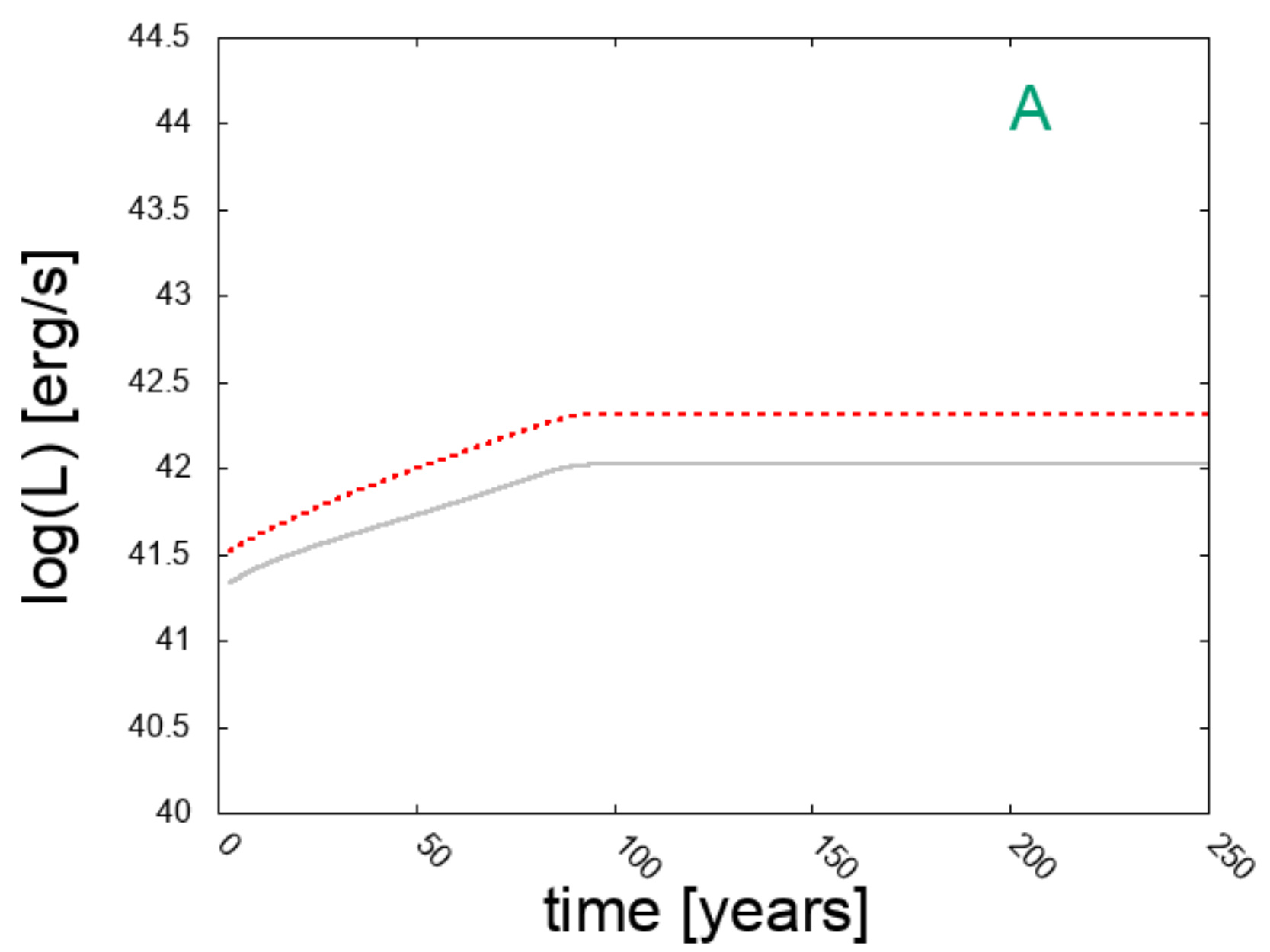}\hspace{-0.2cm}
\includegraphics[scale=0.32]{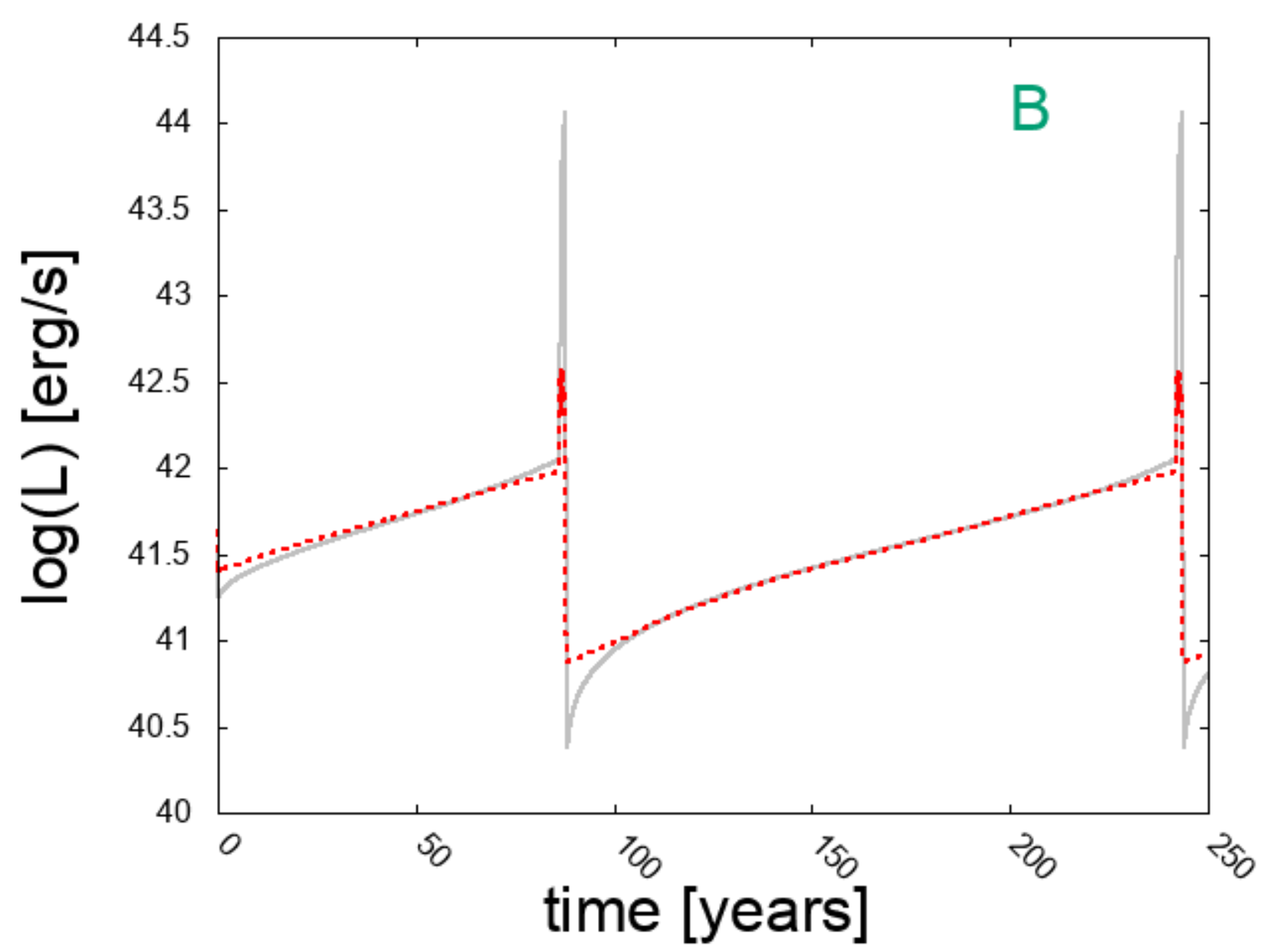}\hspace{-0.2cm}
\includegraphics[scale=0.32]{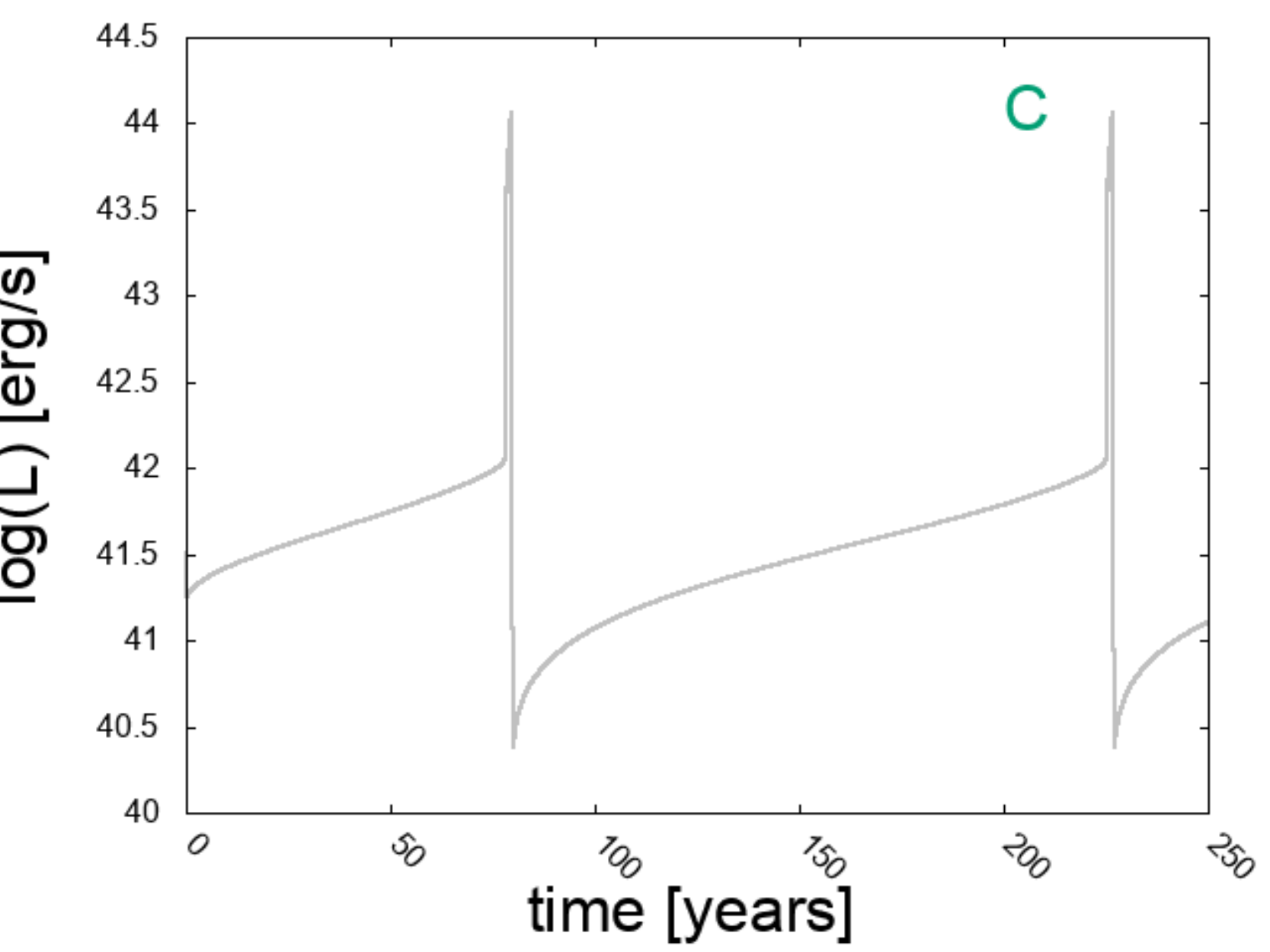}\hspace{-0.2cm}
\caption{Comparison of the modeled variations, for disc (gray line) and corona (red dotted line) luminosity time evolution for 10$^7$M$\odot$ for three scenarios: case A - mass exchange between disk and corona is on and mass evaporation is due to the electron conduction (left panel), case B -  mass exchange between disk and corona is on and mass evaporation is due to magnetic buoyancy (middle panel), case C - mass exchange between disk and corona is off (right panel).} \label{lightcurvemodels}
\end{figure*}

\begin{figure}[t]
\centerline{\includegraphics[scale=0.5]{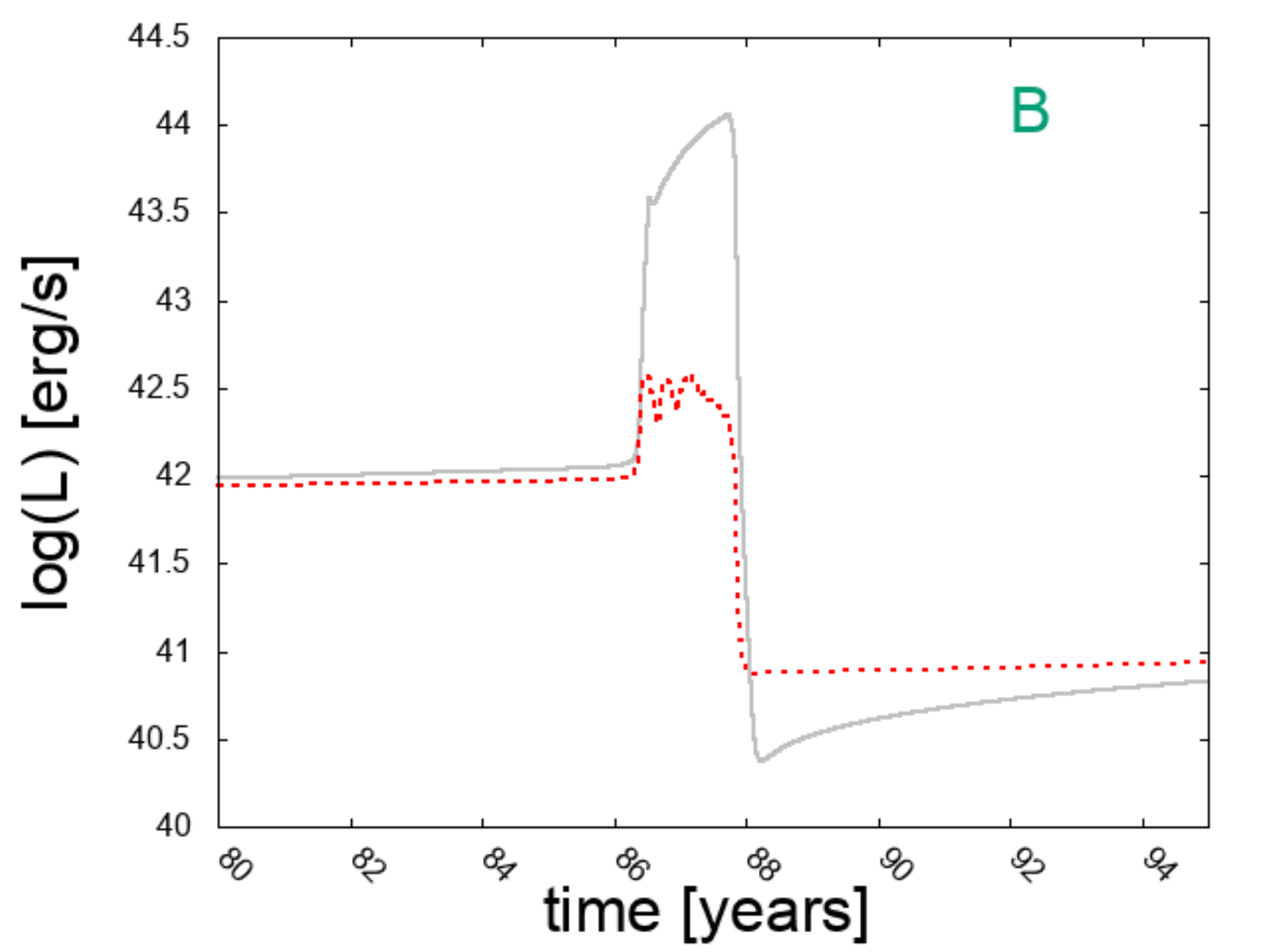}}
\caption{The exemplary fragment of the disk (marked in gray) and corona (marked as dotted red line) lightcurve  for case B.  \label{caseb-zoom}}
\end{figure}

\subsection{Time-dependent solutions}
\label{sect:time_dependent}
In Fig.\ref{lightcurvemodels} we show the modeled lightcurves of the disk (marked in gray) and the corona (marked as red dotted line).
This calculation is representative for objects with observed repetitive outbursts (like NGC  1566, with log(M) = 6.92 $ M_\odot$ \citep{woo2002}).
We fixed the viscosity coefficient in the disk and corona $\alpha$ = 0.02 for both, as we stress out in Section \ref{itemizetimedependent}.
We concentrate on the comparison of the three physical scenarios included in the model. The lightcurve for corona is shown only in cases when disk-corona mass exchange was used in computations.
In the first model (left panel of Fig. \ref{lightcurvemodels} - case A), for assumed physical parameters disk and corona are stabilized. No outbursts are seen, and the system simply evolves from the initial conditions to the equilibrium solution.
For two other investigated scenarios outbursts are present.
The characteristic three phases of the outbursts are thus visible both when the disk/corona coupling is described by magnetic buoyancy
(middle panel of Fig. \ref{lightcurvemodels} - case B) as well as for model with no corona included (right panel of Fig. \ref{lightcurvemodels} - case C). The diffusive phase is the longest of the three phases. 

The outbursts in both cases are very strong, with the amplitude (the ratio of the maximum to the minimum flux in the disk $\frac{L_{max}}{L_{min}}$) $\approx 10^4$. The period of the outburst for the case B is 156 years.
Flare duration to period ratio \footnotetext{As a flare duration we assume the time between $(L_{max}+L_{min})/2$ approached during the rise of the outburst, and  $(L_{max}+L_{min})/2$ approached during the dimming, following parametrization from \cite{grzedzielski2017}.} is 0.006. 

For corona in case B (see Fig.~\ref{caseb-zoom} for the expanded illustration of both disk and corona outburst) the luminosity starts to increase $\approx 0.5$ year before the disk's outburst and gets to the lowest luminosity level $\approx 0.5$ year before the disk's dimming, so the period of the outburst is the same for corona and accretion disk. The relative amplitude, in this case, is 50, and the flare duration to period ratio is 0.009.
Thus for these specific model parameters the outbursts should be seen predominantly in the UV band, with more moderate outbursts in X-ray.
In case C the relative amplitude of the outburst is 5000, by a factor 2 smaller than for case B, with the period of 147 years, which is shorter than in case B. Flare duration to period ratio is 0.007.









\section{DISCUSSION}\label{sec4}

We performed simulations of the disk/corona/inner ADAF system (see Fig. \ref{fig:models}) under the radiation pressure instability in order to check if this mechanism 
can explain the CL behavior, as proposed by \cite{2020A&A...641A.167S}. In the preliminary one-zone model of \cite{2020A&A...641A.167S}, the expected timescales were much shorter than 
the local viscous timescale due to the narrowness of the instability zone. In the present computations, the outer radius is set at 100 $R_{Schw}$, and the extension of the instability zone is determined numerically. During the outburst, the larger part of the disk becomes a subject of instability than just the narrow unstable zone as described by the 
stationary S-curves (see e.g. \citealt{grzedzielski2017}). Therefore, the timescales cannot be arbitrarily shortened by selecting the specific accretion rate and the 
transition radius between the disk/corona part and the inner ADAF. Here we show only one example for the global parameter values (black hole mass $10^7 M_{\odot}$, $\dot m = 0.007$) but 
the same holds for other values of the black hole and accretion rates (Sniegowska et al., in preparation). 

We show that the model is actually sensitive to various assumptions about the role of the electron conduction and/or magnetic field inside the disk, and the description of the mass exchange between the disk
and the corona. The disk is stabilized in the model A. The outbursts are slightly shorter if the mass exchange between the disk and the corona is turned off (see Fig.~\ref{lightcurvemodels}). 
The duration of the outburst in the presented model is very 
short in comparison with the separation of outbursts. 


The timescales derived in Sect.~\ref{sect:time_dependent} are the order of 150 years, shorter than the outbursts timescales obtained 
by \cite{grzedzielski2017} in case of modified viscosity (using the maximum to the minimum luminosity in the outburst from 
Fig.~\ref{lightcurvemodels} and using with Eq.~31 of \citealt{grzedzielski2017} we obtain 7000 years). This effect is not related to the presence of the corona which also transports a 
fraction of the accreting material, but most of the effect is due to the small outer disk radius used in current computations. 
Radiation pressure instabilities mechanism would lead  to outbursts of the much larger part of the disk than 
100$R_{Schw}$ (which we consider in this work as the outer radius of simulated accretion disk). We checked that by computing one more model for case B, adopting the outer radius at 80$R_{Schw}$, and those computations gave a total period of 85 years, and apparently, the outer radius is the most critical parameter of the current model.

This is a very interesting aspect of the model, but we currently cannot address properly it without considerable modification of the GLADIS code. Currently, we assume that the outer disk is stationary. If the disk outer radius is indeed very small since the
event is related to TDE, we would need to model the whole disk as initially a ring, with a freely expanding outer radius due to the viscous evolution. We will address this in the future.



Those results are not yet ready for detailed comparison with the lightcurve for observed AGN, because the calculated zone is not 
extended enough to do so. In the incoming paper (Sniegowska et al., in preparation) we will study a much broader parameter range. However, they show the critical role of the outer disk radius, and, on the other hand, a moderate dependence
of the modifications of the description of the disk structure. This second aspect is very promising for the future, since the 
disk structure, corona formation and the transition to ADAF are still subject of vigorous modeling, both numerical and analytical 
or semi-analytical.

Summarizing, we see from the simulations presented in this paper that the timescales of the order of years can be recovered from the model, after proper adjustment of the parameters. However, rapid oscillations discovered recently in two sources and named the Quasi-Periodic Ejections seem to be beyond the reach of the radiation pressure instability model. Below we include a short discussion of these observed phenomena.


\subsection{Case study - timescales in NGC 1566}
NGC 1566 is one of the sources in which outbursts seem to appear semi-repetitively, in the timescales of years, it is thus a source which may be modeled by radiation pressure instability, as postulated by \citet{2020A&A...641A.167S}. 
First observations of the source which report optical variability are obtained by \cite{1975MNRAS.173P..57Q}. 
Later, changes in H$\beta$ and H$\alpha$ lines were detected, as well as the change in spectral continuum \cite{1985ApJ...288..205A}. Monitoring of those changes was continued by \cite{1986ApJ...308...23A}.

New changes (hard X-ray emission) in this source were detected
by \cite{2018ATel11754....1D} with INTEGRAL \citep{2003A&A...411L...1W} mission
and then followed up with using SWIFT observatory data by \cite{2018ATel11783....1F}.
Those observations were complemented by photometry and spectroscopy later on.
\cite{2019MNRAS.483..558O} report the increase in the luminosity by 25-30 times in June 2018, which was followed by smaller outbursts and makes this source even more interesting.

\cite{2021MNRAS.507..687J} analyse \textit{XMM-Newton} \citep{2001A&A...365L...1J}, \textit{NuSTAR} \citep{2013ApJ...770..103H} and \textit{Swift} \citep{10.1111/j.1365-2966.2009.14913.x} X-ray band archival observations of NGC 1566 from the period between 2015 and 2019.



Amplitudes in the X-ray band change by the factor of 70, whereas bolometric luminosity by the factor of 25. Our models shown in Fig.~\ref{lightcurvemodels} imply higher amplitude in the UV band, but searching the parameter space we can possibly find more suitable solutions. The timescales of the order of a few years are also longer than what we show in Fig.~\ref{lightcurvemodels}, but allowing for higher viscosity parameter, adjusting the parameters in such a way as to get lower outburst amplitude, and (more importantly) for still lower outer radius we could shortened the period. 
Using the timescale for radiation pressure dominated outburst from \cite{grzedzielski2017} for the 10$^7$ solar masses and change in luminosity factor as 25 for this source we get 135 years as the minimal timescale of outburst. However, in this paper modified viscosity was used (which in general shortened the period) but the outer radius was large allowing for unconstrained radial development of the outburst. A smaller outer radius in these models also could easily shorten the period to the requested few years timescale.

\subsection{Case study - timescales in RX J1301.9+2747}

The regular quasi-periodic eruptions were discovered so far in two sources \citep{Miniutti2019,Giustini2021}. The second of these two papers brings the discovery of regular X-ray outbursts in the timescales of the order of 5 hours. The black hole mass in this source is lower than the value used in our simulations ($\sim 10^6 M_{\odot}$ vs. $\sim 10^7 M_{\odot}$, so the timescales from the model should be scaled down by (at least) the mass ratio but such short timescales cannot be reached within the frame of the radiation pressure instability model considered in this paper. As discussed by \citet{Giustini2021}, the underlying mechanism requires a large height to radius ratio, i.e. the instability must happen not in the cold disk zone but (most likely) in the warm corona. In our model, we have only hot corona at local virial temperature which is stable, and the warm corona is not present. Hints for thermal instabilities come from the problems to derive stationary warm corona solutions for some parameter range \citep[see][]{Gronkiewicz2020} but no time-dependent models are available so far.

\section{Summary}
In this work:
\begin{itemize}
    \item we perform the full computations of time-dependent evolution of the accretion disk induced by the radiation pressure using GLADIS code, 
    \item we confirm that it is possible to obtain limit-cycle oscillations for the black hole mass 10$^7$M$_{\odot}$ using this model, 
    \item inclusion of the inner ADAF and coronal flow does not shorten the outburst,
    \item the timescales from the model are not very sensitive to the disk/corona coupling if outbursts are present, but the amplitudes do change, 
    \item the outburst timescales shorten strongly with the decrease of the outer disk radius,
    \item the phenomenon of CL AGN seems to be complex and challenging to model, since observed objects show different behaviors. The radiation pressure model may describe (after further modifications) the yearly variability in the sources like NGC 1566, but most probably a different mechanism must account for Quasi-Periodic Ejections.

\end{itemize}
However, these results require confirmation, with testing the different resolution. The code is grid-based, and such hydrodynamical codes are known to have resolution-dependent convergence.


  

\section*{Acknowledgments}
MS acknowledges the organizers and participants of 13th Serbian Conference on Spectral Line Shapes in Astrophysics for the fruitful discussion and comments which help to improve this work. BC and MS acknowledge the financial support from the Polish Funding Agency National Science Centre, project 2017/26/A/ST9/-00756 (MAESTRO 9) and MNiSW grant DIR/WK/2018/12.







The package (tar.gz) with the source code GLADIS is available at
$http://www.cft.edu.pl/astrofizyka/$ under tab Numerics.

\bibliography{marzena_sniegowska}%

\end{document}